# Anti-measurement Matrix Uncertainty Sparse Signal Recovery for Compressive Sensing

[1,2]Yipeng Liu, [1]Qun Wan, [1]Fei Wen, [2]Jia Xu, [2]Yingning Peng

[1]Department of Electronic Engineering, University of Electronic Science and Technology of China (UESTC), Chengdu 611731, China

[2]Department of Electronic Engineering, Tsinghua University, Beijing, 100084, China

## ABSTRACT

Compressive sensing (CS) is a technique for estimating a sparse signal from the random measurements and the measurement matrix. Traditional sparse signal recovery methods have seriously degeneration with the measurement matrix uncertainty (MMU). Here the MMU is modeled as a bounded additive error. An anti-uncertainty constraint in the form of a mixed $\ell_2$ and $\ell_1$ norm is deduced from the sparse signal model with MMU. Then we combine the sparse constraint with the anti-uncertainty constraint to get an anti-uncertainty sparse signal recovery operator. Numerical simulations demonstrate that the proposed operator has a better reconstructing performance with the MMU than traditional methods.

# 1. INTRODUCTION

Estimation of an unknown sparse vector from a limited number of observations is a common problem. It requires reconstruct the signal with a much fewer randomized samples than Nyquist sampling with high probability on condition that the signal has a sparse representation (Candes and Wakin, 2008). The vector is said to be sparse in that most of the coefficients are zero or much small and only a few are large. The sparse signal recovery refers to the problem of correctly estimating the positions and amplitudes of the non-zero entries from a set of linear measurements. It is of broad interest, arising in various areas, including subset selection in regression, structure estimation in graphical models, sparse approximation, signal denoising.

Most the sparse signal recovery algorithms, such as orthogonal matched pursuit (OMP) (Needell and Vershynin, 2009), basis pursuit (BP) (Chen et al., 1999) and Dantzig Selector (DS) (Candes and Tao, 2007), did not consider the uncertainty in the measurement matrix. In fact the matrix uncertainty is that the measurement matrix is observed with additive error. The measurement matrix uncertainty (MMU) problem arises in many applications including image processing with finite frequency grids, channel estimation with finite channel impulse response (CIR) grids, ADC with aliasing, jitter, finite quantization, aperture effect, and non-linear effects. It is common problem in signal processing. Under MMU, the present sparse signal recovery algorithms (Candes and Tao, 2007; Chen et al., 1999; Needell and Vershynin, 2009) turn out to be extremely unstable. Apart from signal processing, The MMU also arises in many other areas, such as model selection with missing data, portfolio selection, and so on.

To prevent the performance degradation caused by the MMU, This paper proposes a robust sparse signal recovery method. Combining with the sparse constraint in the form of minimization of the $\ell_1$ norm, a mixed $\ell_2$ and $\ell_1$ norms constraint with respect to the MMU parameter is taken to allow certain uncertainty in the measurement matrix for sparse signal recovery. Numerical experiments demonstrate that the proposed robust sparse signal recovery has a better performance than traditional way..

## 2. SIGNAL MODEL

Considering the random measurement model with MMU:

$$\mathbf{y} = \mathbf{A}\boldsymbol{\theta}, \quad \boldsymbol{\theta} \in \mathbf{R}^N, \qquad (1)$$

$$\mathbf{B} = \mathbf{A} + \mathbf{V}, \qquad (2)$$

where $\mathbf{y} \in \mathbb{R}^M$ is the random samples, the real $\theta$ random measurement matrix $\mathbf{A}$ is unknown, $\mathbf{B}$ is the $M \times N$ observed measurement matrix with an additive noise $\mathbf{V}$, and the $N \times 1$ vector $\boldsymbol{\theta}$ is sparse with most of its elements are zero or close to zero.

Without loss of generality, we mainly assume that $\mathbf{V}$ is random but bounded:

$$\|\mathbf{V}\|_\infty \leq \delta, \qquad (3)$$

where $\delta > 0$ and $\|\cdot\|_\infty$ stands for $\ell_\infty$ matrix norm.

## 3. TRIDITIONAL SPARSE SIGNAL RECOVERY

There are two groups of traditional methods to reconstructing the sparse signal. One is the convex programming, such as basis pursuit (BP) (Chen et al., 1999) and Dantzig Selector (DS) (Candes and Tao, 2007), etc; and the other is greedy algorithm, such as Orthogonal matched pursuit (OMP) (Needell and Vershynin, 2009, iterative thresholding (Blumensath and Davies, 2009), etc. BP has almost the same performance as DS (James et al, 2009). The convex programming has better reconstruction accuracy than greedy algorithms; while the greedy algorithm has a less Computational complexity. For a higher accuracy, BP is usually chosen to reconstruct the sparse signal.

To encourage sparsity, The $\ell_0$ optimization is optimal but non-convex and known to be NP-hard. This practice approximation is known as basis pursuit (BP) which is a principle for decomposing a signal into an "optimal" superposition of dictionary elements, where optimal means having the smallest $\ell_1$ norm of coefficients among all such decompositions. It can be formulated with the observed random measurement matrix B and the random measurement vector y as:

$$\min_{\boldsymbol{\theta}} \|\boldsymbol{\theta}\|_1, \quad \text{s.t. } \mathbf{y} = \mathbf{B}\boldsymbol{\theta}. \tag{4}$$

(4) is a second-order cone programming (SOCP). It can be solved by many convex programming tools, such as SeDuMi (Sturm, 1999), etc.

## 3. THE PROPOSED ROBUST SPARSE SIGNAL RECOVERY

As BP is designed without allowing MMU, it mismatches the sparse signal model with MMU (1) (2). The simulation in the next section would also demonstrate that it has a bad performance to estimate the sparse signal **θ** from the measurements **y** and the observed measurement matrix **B**.

To insert the MMU factor in the sparse signal recovery algorithm, we propose an anti- uncertainty constraint (AUC), it can be formulated as

$$\|\mathbf{y} - \mathbf{B}\boldsymbol{\theta}\|_2 \leq \sqrt{M}\delta \|\boldsymbol{\theta}\|_1, \qquad (5)$$

where $\|\mathbf{x}\|_1 = \sum_i |x_i|$ and $\|\mathbf{x}\|_2 = \left(\sum_i |x_i|^2\right)^{1/2}$ are the $\ell_1$ norm and the $\ell_2$ norm of the vector $\mathbf{x} = [x_1, x_2, \ldots, x_N]^T$. It gives the relationship between the square error and sparsity of the estimated signal with the MMU parameter incorporated.

Combining the AUC with the sparse constraint, we get the anti-uncertainty operator (AUO) as:

$$\begin{aligned} & \min \|\boldsymbol{\theta}\|_1 \\ & \text{s.t. } \|\mathbf{y} - \mathbf{B}\boldsymbol{\theta}\|_2 \leq \sqrt{M}\delta \|\boldsymbol{\theta}\|_1 \end{aligned}. \qquad (6)$$

It can be reformulated as

$$\begin{aligned} & \min (t) \\ & \text{s.t. } \|\mathbf{y} - \mathbf{B}\boldsymbol{\theta}\|_2 \leq \sqrt{M}\delta t. \\ & \quad \|\boldsymbol{\theta}\|_1 < t \end{aligned} \qquad (7)$$

This is a convex programming, and can be solved by software as SeDuMi (Sturm, 1999). With a tighter constraint on the squared error and the sparsity, the denoising performance would be improved, and higher reconstruction accuracy would be obtained.

The detailed derivation of AUC is as follows:

$$\begin{aligned}\|\mathbf{y}-\mathbf{B}\boldsymbol{\theta}\|_2 &= \|\mathbf{y}-(\mathbf{A}+\mathbf{V})\boldsymbol{\theta}\|_2 \\ &= \|\mathbf{A}\mathbf{r}-(\mathbf{A}+\mathbf{V})\boldsymbol{\theta}\|_2 \\ &= \|\mathbf{V}\boldsymbol{\theta}\|_2\end{aligned} \qquad (8)$$

Here we define the row vector $\mathbf{v}_m$, m = 1, 2, … , $M$ of the matrix $\mathbf{V}$ as:

$$\mathbf{V} = \begin{bmatrix} \mathbf{v}_1 \\ \mathbf{v}_2 \\ \vdots \\ \mathbf{v}_M \end{bmatrix}. \qquad (9)$$

Then (8) can be reformulated as:

$$\|\mathbf{V}\boldsymbol{\theta}\|_2 = \sqrt{\sum_{m=1}^{M}|\mathbf{v}_m\boldsymbol{\theta}|^2}, \qquad (10)$$

where $|\bullet|$ means the modulus operator of a scalar. It is easy to prove that

$$\begin{aligned}|\mathbf{v}_m\boldsymbol{\theta}| &= |v_{m,1}\theta_1| + |v_{m,2}\theta_2| + \cdots + |v_{m,N}\theta_N| \\ &\leq (|v_{m,1}|+|v_{m,2}|+\cdots+|v_{m,N}|)(|\theta_1|+|\theta_2|+\cdots+|\theta_N|), \quad (11) \\ &= \|\mathbf{v}_m\|_1 \|\boldsymbol{\theta}\|_1\end{aligned}$$

where $v_{m,i}$, $i = 1, 2, \ldots, N$, is the i-*th* element of the vector $v_m$; and $\theta_i$, $i = 1, 2, \ldots, N$, is the i-*th* element of the vector $\boldsymbol{\theta}$. Taking (11) into (10), we can get

$$\|\mathbf{V}\boldsymbol{\theta}\|_2 \leq \sqrt{\sum_{m=1}^{M}\left(\|\mathbf{v}_m\|_1 \|\boldsymbol{\theta}\|_1\right)^2} . \qquad (12)$$

Then, obviously we have

$$\begin{aligned}\|\mathbf{V}\boldsymbol{\theta}\|_2 &\leq \sqrt{\sum_{m=1}^{M}\left(\|\mathbf{v}\|_{1,\max} \|\boldsymbol{\theta}\|_1\right)^2} \\ &= \sqrt{M\left(\|\mathbf{v}\|_{1,\max} \|\boldsymbol{\theta}\|_1\right)^2} , \qquad (13) \\ &= \sqrt{M}\, \|\mathbf{V}\|_\infty \|\boldsymbol{\theta}\|_1 \end{aligned}$$

where $\|\mathbf{v}\|_{1,\max}$ means the maximum one of all the values of the $\|\mathbf{v}_i\|_1$, $i = 1, 2, \ldots, M$. With the condition (3), we can get:

$$\|\mathbf{V}\mathbf{r}\|_2 \leq \sqrt{M}\,\delta \|\mathbf{r}\|_1 . \qquad (14)$$

Combining (8) and (14), we get the AUC (5).

It is a relaxing version of the standard square error bound constraint and gives the relationship between the square error and sparsity measure of the estimated signal with the sampling distortion parameter incorporated.

## 4. SIMULATION

In this section we present simulation results to demonstrate the performance gain of the AUO. Without loss of generality, we assume the length of the sparse signal is $N = 500$, the length of random measurement vector is $M = 125$ and the number of nonzero entries of the sparse signal is $K = 6$. The positions of the nonzero entries are randomly distributed, and the amplitudes of the nonzero entries are all set to be one. **V** is a Gaussian random matrix with the bound of the measurement matrix elements being $\delta = 0.7$. The real measurement matrix **A** is the random Sub-sampling matrices which are generated by choosing $M$ separate rows uniformly at random from the unit matrix.

Figure 1 demonstrates that the real sparse signal and the normalized sparse signals recovered by the BP and the AUO with 1000 independent trials were averaged. Comparing with the real sparse signal, it shows that the noise level is too high to correctly distinguish the nonzero entries of the sparse signal recovered by the BP. However the AUO can successfully suppress the noise. The recovered signal gives out conspicuous nonzero entries with the positions correctly corresponding to the real sparse signal.

To further demonstrate the performance of the proposed AUO, Monte Carlo simulation is used here. A new performance evaluation function is defined to represent the number of incorrectly estimated elements in $L$ Monte Carlo simulation:

$$\rho = \frac{\sum_i \#_i(P_0 | P_1) + \#_i(P_1 | P_0)}{L} \tag{15}$$

where $\#_i(P_0 | P_1)$ counts the number of estimated nonzero elements which should be zero in fact; and $\#_i(P_1 | P_0)$ counts the number of estimated zero elements which should be nonzero in fact.

Figure 2 gives the performance function $\rho$ with different number of nonzero elements. Obviously it shows that both AUO and BP have their $\rho$ increase with the increase of the number of nonzero elements. Besides, it demonstrates that the AUO outperforms BP for it achieves smaller number of incorrectly estimated elements with the number of nonzero elements. Figure 3 gives the performance function $\rho$ with different number of measurement numbers. It shows that both AUO and BP have their $\rho$ decrease with the increase of the number of measurements. Besides, it can be seen that with the same number of measurements, AUO achieves smaller $\rho$ than BP. In the condition of this simulation, AUO's the performance gain against BP is more obvious from $M = 30$ to $M = 140$.

## 5. CONCLUSION

In this paper, we propose a robust sparse signal recovery with the measurement matrix uncertainty. By Combining the AUC and $\ell_1$ norm minimization, performance gain against BP is obtained when the measurement matrix uncertainty is existed.

In the future, the theoretical performance evaluations can be researched to further solidate the performance of the proposed method. Besides, the proposed AUO is a convex programming. The corresponding greedy algorithm can be developed. Finally, the proposed AUO estimates a single vector from a single snap, and it can be generalized to the Multiple Measurement Vectors (MMV) situation.

## ACKNOWLEDGEMENT

This work was supported in part by the National Natural Science Foundation of China under grant 60772146, the National High Technology Research and Development Program of China (863 Program) under grant 2008AA12Z306, the Key Project of Chinese Ministry of Education under grant 109139, China National Science Foundation under Grant 60971087, China Ministry Research Foundation under Grant 9140A07011810JW0111 and 9140C130510D246, Aerospace Innovation Foundation under Grant CASC200904.

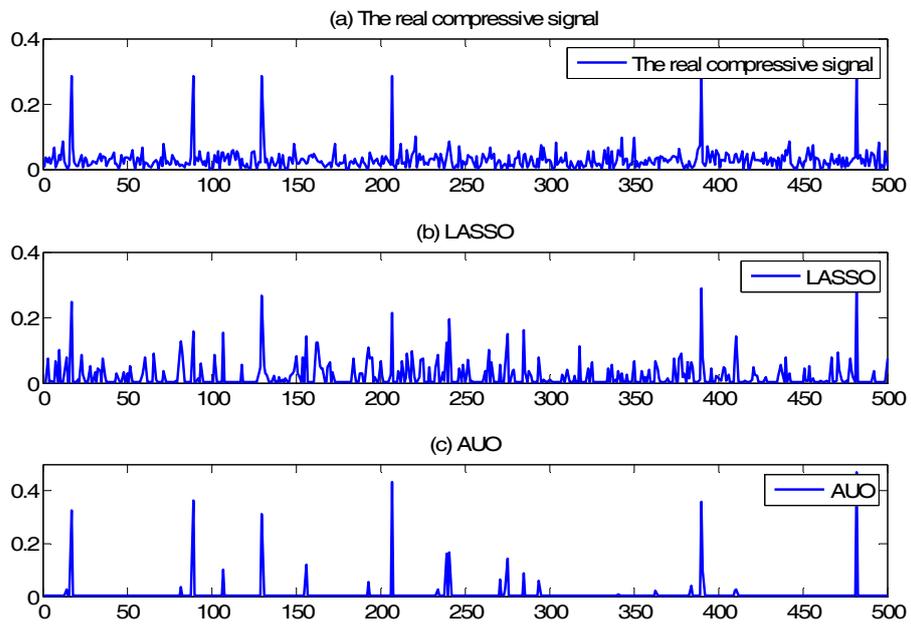

Figure 1  The normalized real sparse signal, the sparse signals recovered by the BP and the AUO with 1000 independent trials were averaged.

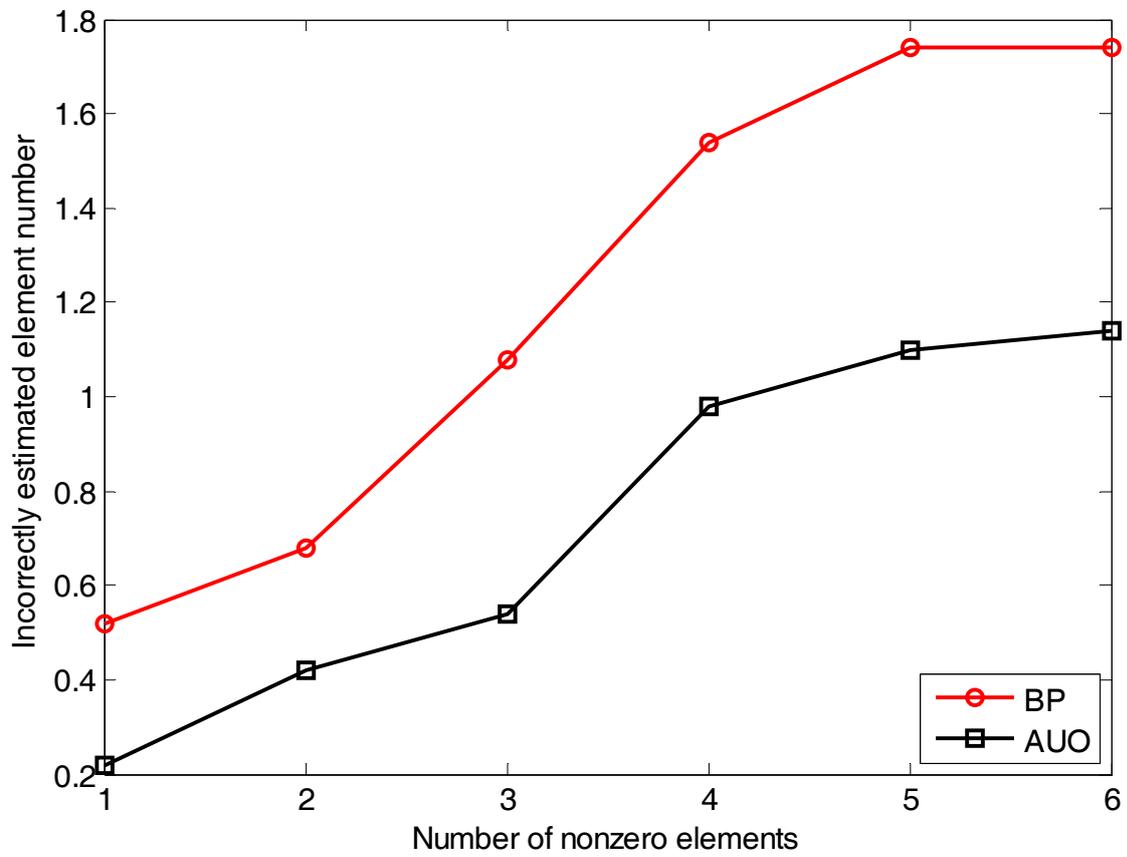

Figure 2  The incorrectly estimated element numbers of AUO and BP with different number of nonzero elements.

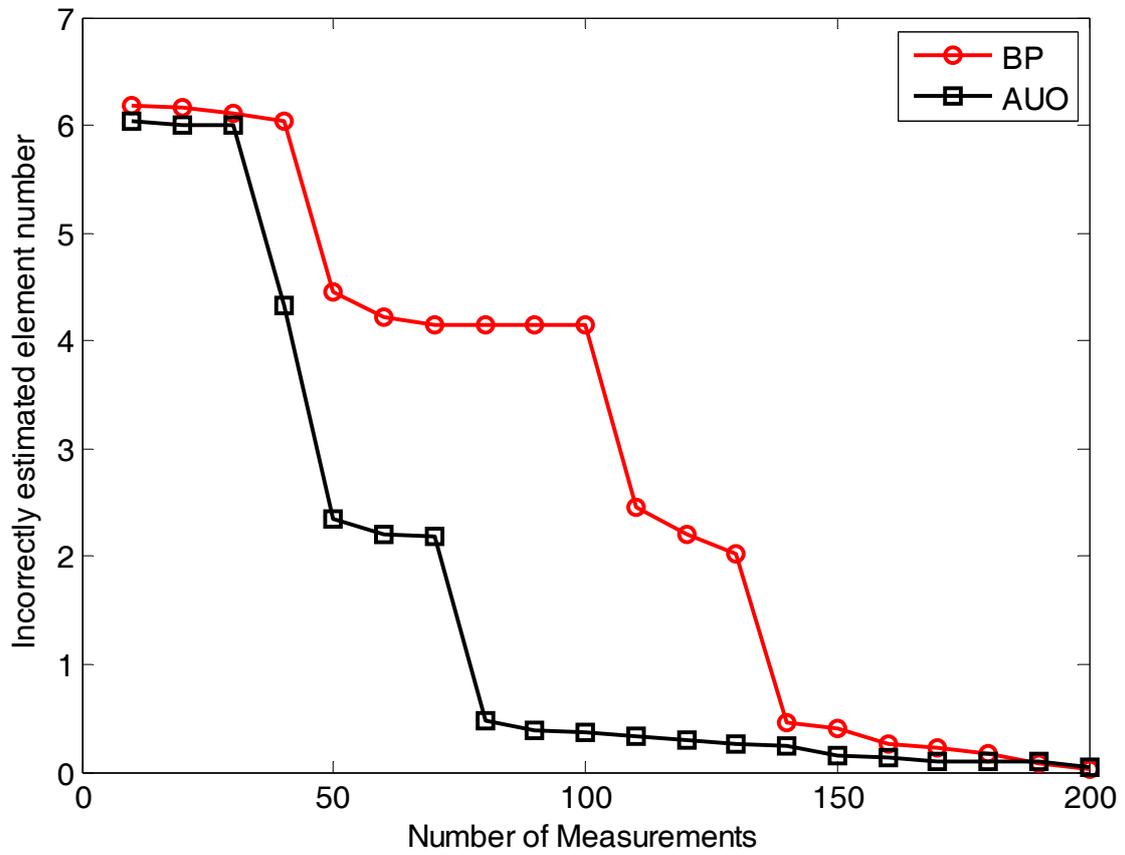

Figure 3  The incorrectly estimated element numbers of AUO and BP with different number of measurements.